\documentclass[fleqn,usenatbib]{mnras}

\usepackage {psfig}
\usepackage{amssymb}
\usepackage {graphicx}
\usepackage {epsfig}
\usepackage{mathrsfs}
\usepackage[english]{babel}
\usepackage{amsfonts}
\usepackage{fontenc}
\usepackage{textcomp}
\usepackage{microtype}
\usepackage[pdftex]{lscape}
\usepackage{amsmath}
\title[Radial-orbit instability with  $r^{-\alpha}$ forces]{Radially anisotropic systems with $r^{-\alpha}$ forces.\\II: radial-orbit instability}
\author[P. F. Di Cintio, L. Ciotti and C. Nipoti]{Pierfrancesco Di Cintio$^{1,2,3}$\thanks{E-mail:
p.dicintio@ifac.cnr.it}, Luca Ciotti
$^{4}$\thanks{E-mail:
luca.ciotti@unibo.it}, and Carlo Nipoti$^{4}$\thanks{E-mail:
carlo.nipoti@unibo.it}\\
$^{1}$Consiglio Nazionale delle Ricerche, Istituto di Fisica Applicata ``Nello Carrara"\\
 via Madonna del piano 10, I-50019 Sesto Fiorentino, Italy\\
$^{2}$Dipartimento di Fisica e Astronomia, Universit\`a di Firenze and Centro Studi Dinamiche Complesse,\\
 via Sansone 1 I-50022 Sesto Fiorentino, Italy\\
$^{3}$INFN -  Sezione di Firenze, via Sansone 1 I-50022 Sesto Fiorentino, Italy\\
$^{4}$Dipartimento di Fisica e Astronomia, Universit\`a di Bologna, viale Berti-Pichat 6/2, I-40127 Bologna, Italy}
\begin{document}
\date{Accepted...  Received...; in original form...}
\pagerange{\pageref{firstpage}--\pageref{lastpage}} \pubyear{0000}
\maketitle
\begin{abstract}
We continue to investigate the dynamics of collisionless systems of particles interacting via additive $r^{-\alpha}$ interparticle forces. Here we focus on the dependence of the radial-orbit instability on the force exponent $\alpha$. By means of direct $N$-body simulations we study the stability of equilibrium radially anisotropic Osipkov-Merritt spherical models with Hernquist density profile and with $1\leq\alpha<3$. We determine, as a function of $\alpha$, the minimum value for stability of the anisotropy radius $r_{as}$ and of the maximum value of the associated stability indicator $\xi_s$. We find that, for decreasing $\alpha$, $r_{as}$ decreases and $\xi_s$ increases, i.e. longer-range forces are more robust against radial-orbit instability. The isotropic systems are found to be stable for all the explored values of $\alpha$. The end products of unstable systems are all markedly triaxial with minor-to-major axial ratio $>0.3$, so they are never flatter than an E7 system.  
\end{abstract}
\section{Introduction}
Self-gravitating collisionless systems with equilibrium initial conditions characterized by a large degree of radial anisotropy (i.e. when a significant fraction of the system's kinetic energy is stored in low angular momentum orbits) are known to be violently unstable. Such instability is commonly referred to as {\it radial-orbit instability}, (hereafter ROI, e.g. see  \citealt{1992JETP...74..755P,2008gady.book.....B,2011TTSP...40..425M,2014dyga.book.....B}).\\
\indent The importance of the ROI for the dynamics of elliptical galaxies is evident. For example, the ROI has been invoked as the physical mechanism responsible for the origin of the tical galaxies riaxiality of ellipt
(see e.g. \citealt{1990ApJ...354...33A,1999A&A...341..361T,2008ApJ...685..739B,2013PhRvL.111w0603P,
2015MNRAS.447...97G,2015MNRAS.446.1335W,2015MNRAS.449.4458S,2015arXiv150503371B}). In the context of the study of the origin of the scaling laws of elliptical galaxies, \cite{2002MNRAS.332..901N} investigated the implications of the ROI on the thinness and tilt of the Fundamental Plane (\citealt{1987ApJ...313...59D,1987ApJ...313...42D}). Remarkably, $N$-body simulations, confirming a conjecture proposed by \cite{1997A&A...321..724C}, revealed that the whole tilt of the Fundamental Plane can not be explained invoking a systematic increase of radial-orbital anisotropy with mass, due to the limits imposed by stability requirements. We finally note that, in Newtonian gravity, numerical simulations show that the presence of a spherical Dark Matter halo has a mild stabilizing effect against ROI (\citealt{1991ApJ...382..466S,1997ApJ...490..136M,2002MNRAS.332..901N}).\\
\indent Notwithstanding the great importance in astrophysics, and the vast amount of work done on this subject (e.g., see \citealt{1981SvA....25..533P,1985MNRAS.217..787M,1991MNRAS.248..494S,1987MNRAS.224.1043P,1994ApJ...434...94B,1996MNRAS.280..689P,1996ApJ...456..274C,
1997ApJ...490..136M,2010MNRAS.405.2785M,2015MNRAS.451..601P}), a fully satisfactory understanding of the ROI has not been reached yet, even though some result is now well established. In fact, the degree of anisotropy of a spherical model can be quantified by the so-called Fridman-Polyachenko-Shukhman stability indicator (\citealt{1984sv...bookR....F})
\begin{equation}\label{index}
\xi=\frac{2K_r}{K_t},
\end{equation}
where $K_r$ and $K_t=K_\theta+K_\phi$ are the radial and tangential components of the kinetic energy tensor, respectively. The main result is that, albeit with some dependence on the specific equilibrium model, for $\xi>\xi_{s}\simeq 1.5\pm 0.2$ the systems are unstable. Unfortunately, it is not clear how much the ROI depends on the density profile of the initial equilibrium configuration, and, at fixed $\xi$, on the anisotropy profile.\\
\indent Of course, even less is known in the case of alternative theories of gravity where the force differs from the Newtonian $1/r^2$ law. Among others we recall the Modified Newtonian Dynamics (MOND, see e.g. \citealt{1983ApJ...270..365M}, \citealt{1984ApJ...286....7B}), the Modified Gravity (MOG, see e.g. \citealt{2006JCAP...03..004M,2013MNRAS.436.1439M}), the Emergent Gravity (see e.g. \citealt{2016arXiv161102269V}), and the so-called $f(R)$ gravities (see e.g. \citealt{1970MNRAS.150....1B,RevModPhys.82.451,2011PhRvD..83d4007Z}).\\
\indent In the context of MOND, for example, it is natural to ask whether radially anisotropic MOND systems are more or less prone to the ROI than their equivalent Newtonian systems\footnote{The ENS of a MOND system is a Newtonian system where the baryonic component has the same phase-space distribution as the parent MOND model. This requires the presence of a dark matter halo (that in general is not guaranteed to have everywhere positive density) in the ENS. Similarly, we could introduce the concept of equivalent $r^{-\alpha}$ system, and study the properties of the associated dark matter halo.} (ENSs). \citet[][hereafter NCL11]{2011MNRAS.414.3298N}  found that, on one hand, MOND systems are always more likely to undergo ROI than their ENSs. On the other hand, MOND systems are able to support a larger amount of kinetic energy stored in radial-orbits than one-component Newtonian systems with the same barionic (i.e. total) density distribution.\\
\indent Here we extend the study of the ROI to the case of a family of radially anisotropic \cite{1990ApJ...356..359H} models with additive interparticle $r^{-\alpha}$ forces with $1\leq\alpha< 3$. In this investigation we limit for simplicity to the force exponents in the range $1\leq\alpha<3$. In fact, this range spans the relevant cases of forces with exponent larger and smaller than 2, and also corresponding to the MOND-like case $(\alpha=1)$. We also note that a scale free force law is also expected in the weak field limit of some of the theories mentioned above (see e.g. \citealt{2004PhLA..326..292C,2017arXiv170203430C,PhysRevD.74.107101}).\\
\indent Note that the study of attractive $r^{-\alpha}$ interparticle forces is not new, as it can be traced back to Newton's {\it Principia} (e.g., see \citealt{1995newt.book.....C}). Relatively recently, \cite{2001PhRvE..64e6103I,2001PhRvL..87u0601I} studied the dynamical phase transitions in systems with $r^{-\alpha}$ interactions undergoing violent relaxation (\citealt{1967MNRAS.136..101L}), and \cite{2002PhRvE..66e1112I} and \cite{2012CEJPh..10..676M} investigated the existence and stability of the stationary states of such systems (see also \citealt{2010PhyA..389.4389B}, \citealt{2014PhR...535....1L}, and references therein), while \cite{2013EPJP..128..128C} developed a kinetic theory for generalized power-law interactions. More recently, \cite{2016arXiv160100064C} and \cite{2017arXiv170101865M} extended the original \cite{1941ApJ....94..511C,1943ApJ....97..255C} approach to quantify the dynamical friction force and evaluate collisional relaxation times in Newtonian systems to the case of $r^{-\alpha}$ forces.\\   
\indent \citet[][hereafter DCC11]{2011IJBC...21.2279D} and \cite[][hereafter DCCN13]{2013MNRAS.431.3177D}, studied the collisionless relaxation process and the end products of dissipationless collapses of initially cold and spherical systems of particles interacting via additive $r^{-\alpha}$ forces and characterized by different virial ratios. This approach allowed us to implement a simpler  direct $N-$body code for the simulations of collapses for different force indices $\alpha$ and initial virial ratios, at the cost of relatively longer computational time with respect to particle-mesh MOND simulations (\citealt{2009MSAIS..13...89L}). The present study presents additional technical problems, because in the set-up the initial conditions we must recover the phase-space distribution function and check for its positivity, as a function of the anisotropy radius and force exponent $\alpha$. The analytical set-up of the initial conditions is presented in \citet[][hereafter DCCN15]{PLA:9846257}.\\
\indent The paper is structured as follows. In Section 2 we describe the set-up of the initial conditions and introduce the quantities that we will use to check the stability of numerical models. In Section 3 we show the evolution of isotropic and marginally consistent radially anisotropic systems and study the structural properties of their final states as functions of the force exponent $\alpha$. In Section 4 we determine numerically the critical value of $r_a$ and $\xi$ for stability, and study the evolution and the end products of unstable anisotropic systems. The main results are finally summarized in Section 5. 
\section{Numerical methods}
\subsection{Initial conditions}
In line with previous studies (e.g. \citealt{1997ApJ...490..136M}, \citealt{2002MNRAS.332..901N}, NCL11, DCCN15), we consider the stability of spherical systems with \cite{1990ApJ...356..359H} density profile
\begin{equation}\label{hernquist}
\rho(r)=\frac{Mr_c}{2\pi r(r_c+r)^3},
\end{equation}
where $M$ and $r_c$ are the total mass and the core radius, respectively and
\begin{equation}\label{massprofile}
M(r)=M\times\left(\frac{r}{r_c+r}\right)^{2},
\end{equation}
is the cumulative mass profile.\\
\indent In order to build initial conditions with a tunable degree of radial anisotropy, as required by our experiments, we adopt the standard Osipkov-Merritt (\citealt{1979SvAL....5...42O}, \citealt{1985AJ.....90.1027M}, hereafter OM) parametrization. It is easy to show that the integral inversion needed to recover the phase space distribution function (\citealt{1916MNRAS..76..572E,2008gady.book.....B}) is independent of the force law (while of course the potential is not); we already adopted the OM inversion to set up initial conditions in MOND (NCL11), and for $r^{-\alpha}$ forces (DCCN15). The anisotropic OM distribution function $f(Q)$ for 
our systems is given by
\begin{equation}\label{OM}
f(Q)=\frac{1}{\sqrt{8}\pi^2}\int_Q^{Q_{\rm sup}}\frac{{\rm d}^2\rho_a}{{\rm d}\Phi^2}\frac{{\rm d}\Phi}{\sqrt{\Phi-Q}},
\end{equation}
where
\begin{equation}
Q=E+\frac{J^2}{2r_a^2}.
\end{equation}
$E$ and $J$ are the particle's energy and angular momentum per unit mass, $r_a$ is the anisotropy radius, and the augmented density is defined by 
\begin{equation}\label{augmented}
\rho_a(r)=\left(1+\frac{r^2}{r_a^2}\right)\rho(r);
\end{equation}
note that we do not adopt the convention, common in Newtonian gravity, of using the relative potential and energy. We recall that the velocity-dispersion tensor is nearly isotropic inside $r_a$, and more and more radially anisotropic for increasing $r$. Therefore, small values of $r_a$ correspond to more radially anisotropic systems, and to larger values of $\xi$.\\
\indent For $\alpha-$forces the potential $\Phi$, due to the spherically symmetric density\footnote{The general expressions of $\Phi$ for a generic distribution $\rho(\mathbf{x})$ is given in DCCN15 (eqs. 2.4-2.7; see also DCC11). For the density distribution in eq. (2), when $1<\alpha<3$ it is possible to fix $\mathbf{x}_0=\infty$ and $\Phi(\mathbf{x}_0)=0$, while for $\alpha=1$ we set $\mathbf{x}_0=0$ and $\Phi(\mathbf{x_0})=GM\times\left[1+\ln(r_c/r_n)\right]$, therefore obtaining the expressions (7) and (8). We recall that for $1<\alpha<3$ $\Phi(0)=GM/r_c^{\alpha-1}\times 2{\rm B}(3-\alpha,\alpha)/(\alpha-1)$, where ${\rm B}(x,y)$ is the complete Euler beta function, (DCCN15).} $\rho(r)$, is given by
\begin{equation}\label{phialfa}
\Phi(r)=-\frac{2\pi G}{r}\int_{0}^{\infty}\rho(r^\prime)\frac{(r+r^\prime)^{3-\alpha}-|r-r^\prime|^{3-\alpha}}{(\alpha-1)(3-\alpha)}r^\prime{\rm d} r^\prime
\end{equation}
for $1<\alpha<3$, and 
\begin{equation}\label{phi1}
\begin{split}
\Phi(r)=\frac{\pi G}{r}\int_{0}^{\infty}\rho(r^\prime)\\
\left[(r+r^\prime)^2\ln\left(\frac{r+r^\prime}{r_n}\right)-(r-r^\prime)^2\ln\left(\frac{|r-r^\prime|}{r_n}\right)-2rr^\prime\right]r^\prime{\rm d}r^\prime
\end{split}
\end{equation}
for $\alpha=1$ (DCC11), where $G$ is a dimensional coupling constant and $r_n$ is the scale length used to normalize interparticle distances. In our case,  the natural choice is to use $r_n=r_c$.
\begin{figure*}
\includegraphics[width=0.7\textwidth]{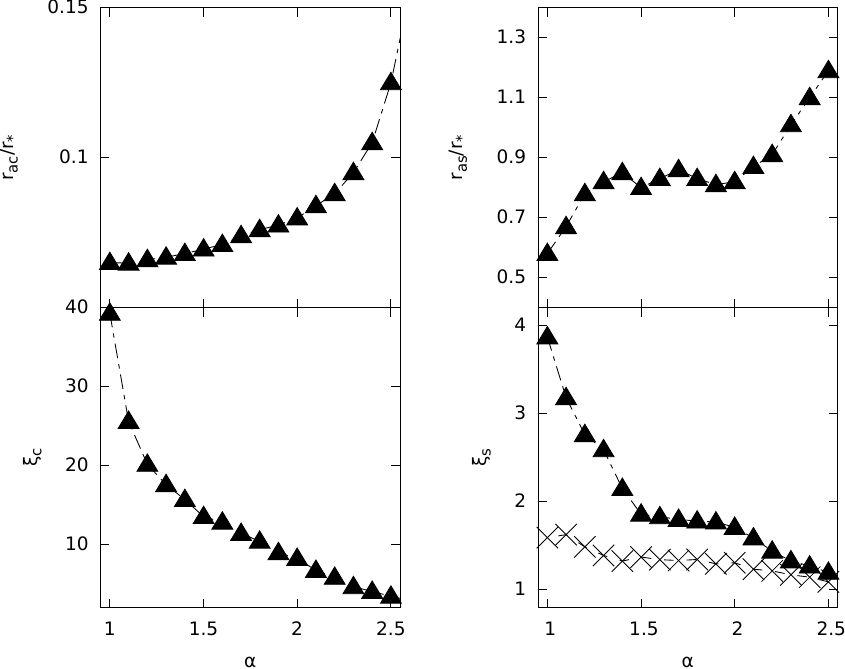}
\caption{Left panels: critical Osipkov-Merritt anisotropy radius for consistency of the Hernquist models, $r_{ac}$ (top, in units of the half mass radius $r_*$) and corresponding value of the stability indicator $\xi_c$ (bottom) as  functions of the force exponent $\alpha$. Right panels: Critical Osipkov-Merritt anisotropy radius $r_{as}$ for stability (top) and corresponding value of $\xi_{s}$ (bottom) as functions of the force exponent $\alpha$. Crosses show the critical value of $\xi$ computed for particles within the half-mass radius $r_*$ only.}
\label{figxistab}
\end{figure*}
From Eqs. (\ref{phialfa}-\ref{phi1}) it is not difficult to show that in eq. (\ref{OM}) $Q_{\rm sup}=0$ for $\alpha>1$ and $Q_{\rm sup}=\infty$ for $\alpha=1$.\\
\indent Since the maximum amount of radial anisotropy that a given system can sustain is limited by the phase-space consistency (i.e. positivity of the phase-space distribution, \citealt{1992MNRAS.255..561C,1996ApJ...471...68C,1999ApJ...520..574C,2009MNRAS.393..179C,2010MNRAS.401.1091C,2010MNRAS.408.1070C,2011ApJ...736..151A,2012MNRAS.422..652A}), it is natural to investigate preliminarily how this limit depends on the force exponent $\alpha$. In fact in DCCN15 we determined, for the family of OM anisotropic Hernquist models, the minimum value $r_{ac}$ and the associated maximum value of the stability indicator, $\xi_c$, for phase-space consistency. In Fig. \ref{figxistab} (left panels) we show these quantities, and their numerical values are given in Tab. \ref{tab1} (columns 2 and 3).\\
\indent As described in in DCCN15, it turns out that $r_{ac}$ increases with $\alpha$, and $\xi_c$ correspondingly decreases, i.e., high values of $\alpha$ (``short-range" forces) lead to phase-space inconsistency in more isotropic models than low values of $\alpha$ (``long-range" forces). In practice, for given $\alpha$, only models with $r_a$ above $r_{ac}$ and $\xi$ below $\xi_c$ are characterized by a nowhere negative $f(Q)$. Therefore, at fixed density profile, systems with lower $\alpha$ can sustain a larger amount of radial OM anisotropy.\\
\indent Once the phase-space distribution function $f(Q)$ is computed, the radial coordinate for the $N$ particles of mass $m=M/N$ is extracted from eq. (\ref{massprofile}). The angular coordinates $\vartheta$ and $\varphi$ are randomly assigned to each particle sampling $\cos\vartheta$ from a uniform distribution in (-1,1), and $\varphi$ from a uniform distribution in $(0,2\pi)$. Then, the potential $\Phi(r)$ is numerically evaluated from Eqs. (\ref{phialfa})-(\ref{phi1}) at the position of each particle. Following \cite{2002MNRAS.332..901N} and DCCN15, we construct the dimensionless vector ${\mathbf u}=(u_1,u_2,u_3)$, with components sampled from a uniform distribution in (-1,1), rejecting the triplets with $||\mathbf{u}||>1$. The putative physical velocity components are then defined as
\begin{equation}
v_r=\sqrt{2|\Phi|}u_1,\quad v_\vartheta=\frac{\sqrt{2|\Phi|}u_2}{\sqrt{1+r^2/r_a^2}},\quad v_\varphi=\frac{\sqrt{2|\Phi|}u_3}{\sqrt{1+r^2/r_a^2}}.
\end{equation}
At this point, eq. (\ref{OM}) is evaluated with $Q=(1-u^2)\Phi$. Applying the von Neumann rejection method, if $f(Q)\geq f(\Phi)$, then the velocity vector $(v_r,v_\vartheta,v_\varphi)$ is accepted. Otherwise, the velocity is discarded and the procedure repeats.
\subsection{The numerical code}
For the numerical simulations we use our direct $N-$body code, already tested and employed to simulate dissipationless collapses of cold systems interacting with $r^{-\alpha}$ forces (DCCN13). In order to compare the results obtained for different values of $\alpha$, we define the time-scale $t_*$ from the dimensionless ratio
\begin{equation}\label{norm}
\frac{GMt_{*}^2}{2r_{*}^{\alpha+1}}= 1,
\end{equation}
where the normalization length scale $r_{*}\equiv (1+\sqrt{2})r_c$ is the half-mass radius of the initial density distribution. The natural velocity scale is then obtained as  
\begin{equation}\label{vnorma}
v_*\equiv\frac{r_*}{t_*}.
\end{equation}
The physical scales $r_*$ and  $t_*$ are used to recast the equations of motion in dimensionless form (DCCN13), which are integrated with a standard second order symplectic integrator (see also e.g., \citealt{doi:10.1080/08927029108022142}). The fixed timestep $\Delta t$ ranges from  $\simeq t_*/80$ for $\alpha=1$ to $\simeq t_*/250$ for $\alpha=2.9$. The divergence in the force and potential at vanishing interparticle separation is prevented by introducing the softening length $\epsilon$, so that $r\rightarrow\sqrt{r^2+\epsilon^2}$ in the potential. The value of $\epsilon$ is chosen as a function of $\alpha$ so that the softened force on a particle placed at $\simeq 5r_*$ from the centre of mass of the system differs by less than $0.01\%$ from the unsoftened force (see DCCN13 and DCCN15 for a discussion). For the simulations in this work $\epsilon$ increases from $10^{-3}r_c$ for $\alpha=1$ up to $6.5 \times  10^{-2}r_c$ for $\alpha=2.9$. With such combination of parameters, the energy conservation is ensured up to one part in $10^5$ at the end of the simulation.\\
\indent All the simulations use $N=20000$ particles and were run up to $40t_*$ on an Intel\textregistered Xeon E5/Core i7 Unix cluster, each simulation taking roughly 80 hours on a single processor. We performed additional simulations with different number of particles (from $N=15000$ up to 30000) for fixed values of $\alpha$ and $r_a$, finding that the main results are unchanged.\\
\indent In the numerical simulations the perturbation responsible to triggering the instability is the numerical noise produced by discreteness effects in the initial conditions plus the round-off error in the orbit integration.
\subsection{Diagnostics}
As indicators of the instability and subsequent relaxation of the models, we monitor the time evolution of the stability indicator $\xi$, of the virial ratio $2K/|W|$ (where $K$ is the total kinetic energy, and $W$ the virial function), of the minimum-to-maximum axial ratio $c/a$, and of the Lagrangian radii $r_5$, $r_{50}$ and $r_{90}$ (i.e. radii enclosing $5\%$, $50\%$ and $90\%$ of the total mass $M$, respectively).\\
\indent As clear from eq. (\ref{index}), a proper definition of $\xi$ can be given only for spherical systems. However, in case of instability spherical symmetry is lost, therefore we need to introduce a more general definition of $\xi$ that can be applied also during the evolution of unstable systems, and that reduces to the standard definition in the spherical case. In order to define the fiducial radial and tangential kinetic energies $K_r$ and $K_t$, that are evaluated at each timestep we proceed as follows, first the position $\mathbf{x}_i$ and velocity $\mathbf{v}_i$ of each particle are referred to the centre of mass of the system as $\mathbf{x}_i^c=\mathbf{x}_i-\mathbf{x}_{\rm cm}$ and  $\mathbf{v}_i^c=\mathbf{v}_i-\mathbf{v}_{\rm cm}$, where $\mathbf{x}_{\rm cm}$ and $\mathbf{v}_{\rm cm}$ are, respectively, the instantaneous position and velocity of the centre of mass as numerically determined by the simulation. Then, for each particle, the radial velocity component is obtained as 
$v_{ri}=\mathbf{v}_i^c\cdot \mathbf{e}_i$, where $\mathbf{e}_i=\mathbf{x}_i^c/||\mathbf{x}_i^c||$ is the radial versor. Therefore, $K_r=\sum_i mv_{ri}^2/2$ and $K_t=K-K_r$.\\
\indent For what concerns the virial ratio, we recall that the virial function $W$ is given by
\begin{equation}\label{viriale1}
W=
\begin{cases}
\displaystyle -\int \rho(\mathbf{x})\langle\mathbf{x},\nabla\Phi\rangle d^3\mathbf{x},\\
\displaystyle \sum_{j\neq i=1}^N m_i\langle\mathbf{x}_i,\mathbf{a}_{ji}\rangle,
\end{cases}
\end{equation}
where the first expression holds for a continuum density $\rho$, and the second for a discrete system of $N$ particles with masses $m_i$ where
\begin{equation}
\mathbf{a}_{ji}=-Gm_j\frac{\mathbf{x}_i-\mathbf{x}_j}{||\mathbf{x}_i-\mathbf{x}_j||^{\alpha+1}}
\end{equation}
is the acceleration at the position of particle $i$ due to particle $j$. For $\alpha\neq 1$ the virial function $W$ is related to the potential energy $U$ of the system (provided $U$ converges for the specific density distribution under consideration) by the identity
\begin{equation}
W=(\alpha-1)U,
\end{equation}
where 
\begin{equation}
U=\frac{1}{2}
\begin{cases}
\displaystyle \int\rho(\mathbf{x})\Phi(\mathbf{x})d^3\mathbf{x},\\
\displaystyle \sum_{i\neq j=1}^N m_i\Phi_{ji},
\end{cases}
\end{equation}
and again the first expression holds in the continuum case, while the second for a system of particles, and $\Phi_{ji}$ is the potential at the position of particle $i$ due to particle $j$.\\ 
\indent Remarkably, for $\alpha=1$ (as for systems in deep-MOND regime, \citealt{2007ApJ...660..256N}, see also \citealt{1992ApJ...397...38G}) $W$ is independent of time, being 
\begin{equation}
W=-\frac{G}{2}
\begin{cases}
\displaystyle M^2,\\ 
\displaystyle \sum_{i\neq j=1}^N m_im_j,
\end{cases}
\end{equation}
and again the first expression holds for a continuum distribution while the second for a system of particles.\\
\indent For what concerns the time evolution of the axial ratio, at given time step the code computes the second order tensor 
\begin{equation}
I_{ij}\equiv \sum_{k} m_k r_i^{(k)}r_j^{(k)},
\end{equation}
where the sum is limited to the particles inside the sphere of Lagrangian radii $r_{90}$, $r_{70}$ and $r_{50}$, (i.e. the radius of the sphere containing 90\%, 70\%  and 50\% of the total mass of the system, respectively). Note that, in terms of $I_{ij}$ the inertia tensor is given by ${\rm Tr}(I_{ij})\delta_{ij}-I_{ij}$. The matrix $I_{ij}$ is iteratively diagonalized, with tolerance set to 0.1\%, to compute the three eigenvalues $I_{11}\geq I_{22}\geq I_{33}$. For a heterogeneous ellipsoid with density stratified over concentric and coaxial ellipsoidal surfaces of semiaxes $a\geq b\geq c$, we would obtain $I_{11}=Aa^2$, $I_{22}=Ab^2$ and $I_{33}=Ac^2$, where $A$ is a constant depending on the density profile. Accordingly, we define the fiducial axial ratios $b/a=\sqrt{I_{22}/I_{11}}$ and $c/a=\sqrt{I_{33}/I_{11}}$, so that the ellipticities in the principal planes are $\epsilon_1=1-\sqrt{I_{22}/I_{11}}$ and $\epsilon_2=1-\sqrt{I_{33}/I_{11}}$. In the following we focus our attention only on the $c/a$ ratio, corresponding to the largest deviation from sphericity.\\ 
\indent Finally, for all simulations, we also consider the evolution of the differential energy distribution $n(E)$, a useful diagnostic in the study of stellar systems (see \citealt{1982MNRAS.201..939V,1982MNRAS.200..951B,1991A&A...249...99C,2006ApJ...637..717T}, DCC11, DCCN13).
$n(E)$ is defined by the relation
\begin{equation}\label{ne}
\int_{E_{{\rm min}}}^{E_{{\rm max}}} n(E){\rm d}E=N,
\end{equation}
where the extremes of integration $E_{{\rm min}}$ and $E_{{\rm max}}$ are the minimum and maximum energies attained by the particles, respectively.\\
\indent The problem of determining the (numerical) stability of a given anisotropic model is a delicate one. As a heuristic criterion, for systems not showing any appreciable evolution of $\xi$ and $2K/|W|$, we have determined first as a function of $\alpha$ the time average $m_{c/a}$ and the standard deviation $\sigma_{c/a}$ of $c/a$ for the {\it isotropic} models over $40t_*$. Typically for the models presented here $m_{c/a}\approx 0.97$ and $\sigma_{c/a}\approx 2.5\times 10^{-3}$, with a very weak dependence on $\alpha$. Then, in order to determine whether a given model is prone to ROI, we monitor the time evolution of its axial ratio $c/a$, and check if its value averaged over the last $20\Delta t$ falls below the value of $m_{c/a}-\sigma_{c/a}$. 
\section{Preliminary experiments}
\subsection{Stability of isotropic models}
Being interested in the stability of anisotropic systems, the first natural question to address is to check whether the isotropic systems are stable for different values of $\alpha$. In case of Newtonian force, analytical stability results are available for the isotropic case, and it is known that phase-space distribution functions with ${\rm d}f(E)/{\rm d}E\leq 0$ correspond to stable systems (the so-called Antonov theorem, see e.g. \citealt{2008gady.book.....B}), moreover, it has been conjectured that models prone to ROI are characterized by non-monotonic $f(Q)$ (e.g., see \citealt{1994ApJ...424..106H}, see also  \citealt{1997ApJ...490..136M,2008gady.book.....B}). When $\alpha\neq 2$, at the best of our knowledge there are not analytical results even in the isotropic case, so we must test the stability of isotropic systems by using $N-$body simulations. Interestingly we found that, for all the explored values of $\alpha$, isotropic Hernquist models are always associated with monotonic distribution functions and are clearly stable. The situation is illustrated in Fig. \ref{figstab} (left panel) where we show, for some representative values of $\alpha$, the evolution of the anisotropy parameter $\xi$, of the virial ratio $2K/|W|$, of the axial ratio $c/a$, and finally, of the Lagrangian radii $r_5$, $r_{50}$ and $r_{90}$. It is apparent that the equilibrium of the initial conditions is preserved for all the considered values of $\alpha$ over the entire simulation up to $t=40t_{*}$. The fluctuations are of the order of $\simeq1\%$ in $\xi$ and $2K/|W|$, of $\simeq5\%$ in $c/a$, and of $\simeq3\%$ in the Lagrangian radii. It follows that the isotropic Hernquist models are numerically stable for $1\leq\alpha<3$. Remarkably, NCL11 found that isotropic MOND systems are also stable.\\
\indent The stability of the isotropic systems is confirmed also by Fig. \ref{figrhoxicrit} (left-hand panel), where we compare the final angle-averaged density profile with the initial profile given by eq. (\ref{hernquist}). The initial density profile is preserved down to $r/r_*\simeq0.025$ in the best case ($\alpha=1.5$) and to $r/r_*\simeq 0.2$ in the worst case ($\alpha=2.5$). The softening length (in units of $r_*$), which fixes the spatial resolution of the simulation is, $\epsilon=2.4\times 10^{-3}$, $1.4\times 10^{-2}$, $2.7\times 10^{-2}$ and $4.5\times 10^{-2}$ for $\alpha=1$, 1.5, 2 and 2.5, respectively.
\subsection{Maximally anisotropic models}\label{s32}
\begin{figure*}
\includegraphics[width=0.8\textwidth]{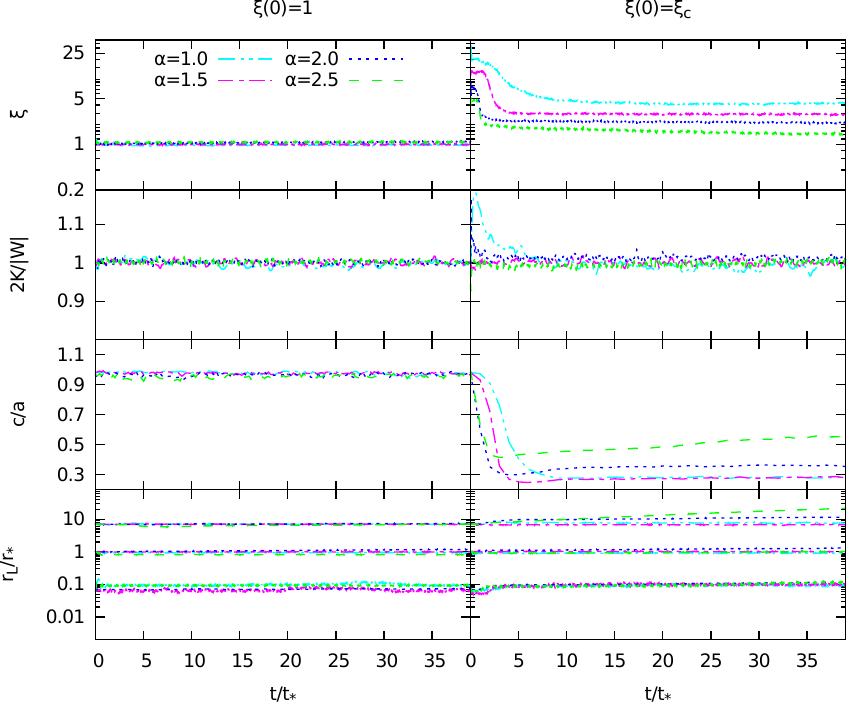}
\caption{From top to bottom: evolution of the stability indicator $\xi$, of the virial ratio $2K/|W|$, of the $c/a$ ratio, and of the Lagrangian radii $r_5$, $r_{50}$ and $r_{90}$ for the isotropic (left panels) and maximally anisotropic (right panels) Hernquist models. Lines corresponding to $\alpha=1,$ 1.5, 2 and 2.5 are indicated with different colours.}
\label{figstab}
\end{figure*}

In a second set of preparatory numerical experiments, we study the evolution of the maximally anisotropic equilibrium models. These marginally consistent models are characterized by the OM phase-space distribution function $f(Q)$ constructed with $r_a\simeq r_{ac}$ (i.e. such that a smaller value of $r_a$ would produce a negative $f(Q)$ for some admissible value of $Q$, see e.g. Fig. (1) in DCCN15), and are therefore associated with the maximum value of the stability indicator $\xi_c$.\\
\indent In Fig. \ref{figstab} (right panel) we show the evolution of $\xi$,  $2K/|W|$, $c/a$ and $r_{5}$, $r_{50}$ and $r_{90}$ up to $40t_*$ for $\alpha=1$, 1.5, 2 and 2.5 and $\xi_c\simeq 39$, 13.5, 8.2 and 3.4, respectively. As expected, the models appear to be violently  unstable. From the evolution of the stability indicator $\xi$ it appears that for all values of the force exponent $\alpha$ the models  rapidly (i.e. within $\simeq t_*$) reduce their degree of anisotropy. The virial ratio $2K/|W|$, with the exception of the $\alpha=1.5$ case, shows a dramatic increase with a peak at $\simeq 0.8t_*$ and then relaxes back to unity within a few $t_*$, a well known feature of violent relaxation. Note however that low amplitude oscillations last over all the simulation time (cfr. the left and right panels of the virial ratio in Fig. \ref{figstab}), and that in general models with a low value of $\alpha$ oscillate for longer times in units of $t_*$ as already found in the MOND case (similar to the $\alpha=1$ model, NCL11), and for simulations of collapses in DCCN13. The reason will be briefly recalled in Section 5.\\
\indent The axial ratio $c/a$ (shown in Fig.\ref{figstab} restricting to particles within $r_{90}$ in order to avoid numerical noise due to a few escapers) also experiences a rapid decrease with the same time-scale (and with the same dependence on $\alpha$) of the other properties. Note that a robust measure of $c/a$ is quite difficult especially in case of high $\alpha$ values: as already found in case of collapses after relaxation such systems produce a characteristic "core-halo" structure with some fraction of their mass (of the order of 5\%) ejected at large distances, but still bound to the main body of the stellar system (DCCN13). In case some of the halo particles belong to the set particles adopted to compute $c/a$ this leads to secular variation of the axial ratio itself. This can be seen in the last panel of Fig. \ref{figstab} where the Lagrangian radii do not present significant evolution except for $r_{90}$ and $\alpha=2.5$, with a slow but systematic increase with time, a consequence of the expansion of the outer regions of the system.\\
\begin{figure*}
\includegraphics[width=0.8\textwidth]{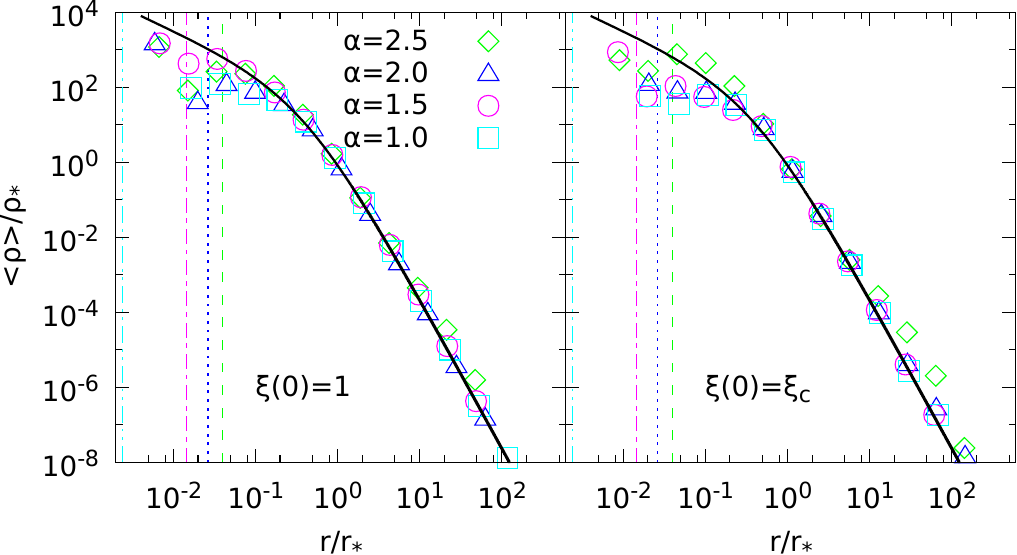}
\caption{Angle-averaged three dimensional (normalized) density profile at the final time $t=40t_*$ for isotropic (left panel) and maximally anisotropic (right panel) Hernquist initial conditions with $\alpha=1$, 1.5, 2 and 2.5. In both panels the heavy solid line represents the analytic Hernquist profile, and the density scale is in units of $\rho_*=\rho(r_*)$ is the density value of the initial condition at the half-mass radius $r_*=(1+\sqrt{2})r_c$. The vertical dashed lines mark the value of the softening length $\epsilon$ used in the simulations.}
\label{figrhoxicrit}
\end{figure*}
\begin{figure*}
\includegraphics[width=0.9\textwidth]{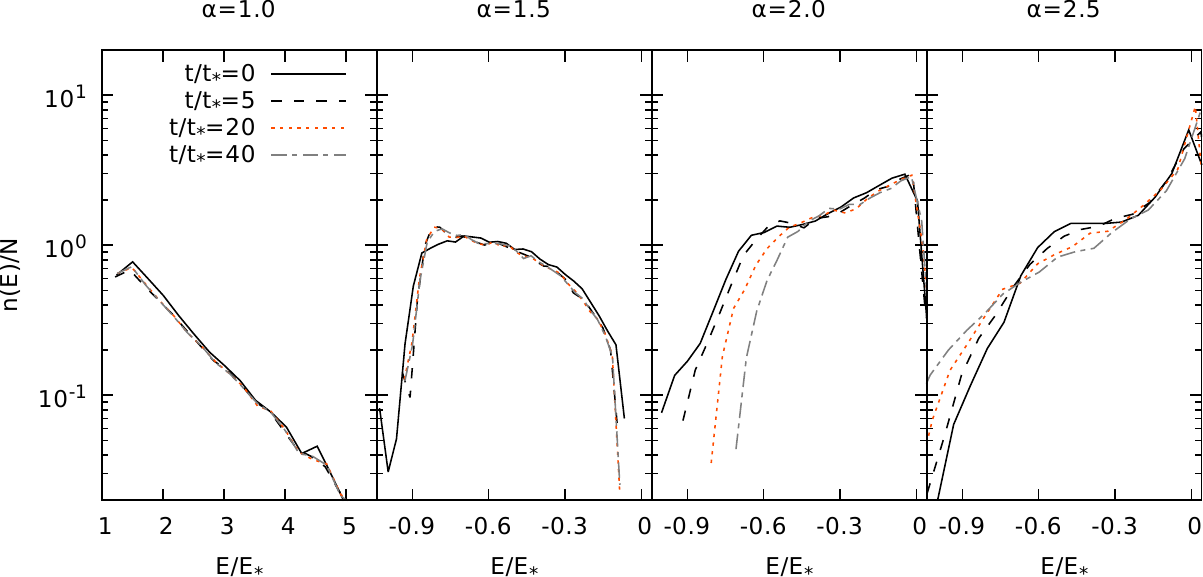}
\caption{Differential energy distribution $n(E)/N$ of the initial conditions (solid lines), and at a selection of times (dashed lines), for maximally anisotropic systems with $\alpha=1.0$, 1.5, 2 and 2.5. For each value of the force index, the normalization energy is $E_*=|\Phi(0)|$ as given in footnote 2.}
\label{fignecrit}
\end{figure*}
\indent The final values of $\xi$ and $c/a$ at $t/t_*=40$ for the maximally anisotropic models for $1\leq\alpha\leq 2.5$ are given in Tab. \ref{tab1} (columns 4 and 5). Note that the end products are less and less anisotropic for increasing $\alpha$, reflecting the trend of the initial conditions. Moreover, the variation of the value of $\xi$, a measure of the redistribution of kinetic energy between radial and tangential motions, decreases for increasing $\alpha$. This trend is also confirmed by a few test simulations (not shown here) with $\alpha$ as large as 2.9. Curiously, at variance with the quite significant dependence on $\alpha$ of the previous quantities,  the final values of the axial ratios are not strongly dependent on $\alpha$ with $(c/a)_{\rm fin}$ ranging from $\simeq0.28$ for $\alpha=1$, to 0.56 for $\alpha=2.5$ (Tab. \ref{tab1}, column 5). The maximum value of the final axial ratio, $\simeq 0.7$ has been obtained for the case $\alpha=2.9$. Therefore, the end products are less flattened for large values of $\alpha$ and, remarkably, no system is found to be significantly flatter than an E7 galaxy\footnote{We note that cold and oblate expanding ellipsoids of charged plasma become at large times prolate with axial ratios limited at $c/a>0.86$. The reasons are however different from instability (see \citealt{2011PhRvE..84e6404G,2014arXiv1408.3857D}).}, similarly to what was found by DCCN13 for the end products of cold collapses with $r^{-\alpha}$ forces.\\
\indent Additional information about the structure of the end products are obtained by inspection of their three dimansional angle averaged density profiles $\langle\rho\rangle$. In Fig. \ref{figrhoxicrit} (right panel) we show $\langle\rho\rangle$ at $t=40t_*$ for the maximally anisotropic initial conditions. Interestingly, when excluding a central region where softening effects are important, the averaged density profiles do not present significant departures from the initial Hernquist density profile (see Fig. 3, right panel solid lines). The only significant feature is associated with the model with $\alpha=2.5$ where a clearly detectable overdensity above the initial profile is present in the range $10<r/r_*<100$: this is a consequence of the already mentioned core-halo structure characteristic of the final configurations of models with high values of $\alpha$, and in fact this feature becomes more prominent for even larger values of $\alpha$ (not shown here).\\ 
\indent In Fig. \ref{fignecrit} we present the evolution of the differential energy distribution $n(E)$ for the maximally anisotropic models with $\alpha=1,$ 1.5, 2 and 2.5. For the low $\alpha$ cases, there is little or no variation in $n(E)$ (dashed lines) with respect to the initial state (solid lines). For larger values of the force exponent, the variations from the initial $n(E)$ are apparent. A few percent of the initial mass in these systems attains a positive energy and escapes after the violent phase in which the ROI takes place (\citealt{2005A&A...433...57T}). Remarkably, the behaviour of $n(E)$ of violently unstable anisotropic models is also strongly resemblant to what found in  cold dissipationless collapses with $r^{-\alpha}$ forces. For example, for low values of $\alpha$, also in DCCN13 the $n(E)$ of the end products was found to depart weakly from that of the initial state.\\
\indent We conclude by noticing that in the context of Newtonian gravity it has been speculated (see e.g. \citealt{2001A&A...378..679E,2005ApJ...634..775B,2006ApJ...653...43M,2007LNP...729..297E}, see also \citealt{1994MNRAS.267..379V}) that $n(E)$ is connected to the form of the {\it circularized} density profile and, if the latter does not change significantly during the process of virialization, $n(E)$ of the final state should not deviate significantly from that of the initial condition. The present findings appear to support this conjecture, as the only models showing significant evolution of $n(E)$ are those with circularized final density profiles that depart more from the initial profile, i.e. those with high $\alpha$.
\section{Results}
\subsection{Critical $r_{as}$ and $\xi_{s}$ for stability}
After having assessed the stability of isotropic and maximally anisotropic systems, we are now in position to determine numerically, as a function of the force exponent $\alpha$, the minimum value of $r_a$ for which the system is stable. We call this critical value $r_{as}$, and once $r_{as}$ is determined we compute the value of the associated stability indicator $\xi_s$. This is one of the main focus of this paper.\\
\indent For each of the 16 different values of the force exponent in the range $1\leq\alpha\leq 2.5$ (see Tab. \ref{tab1}) we performed several runs with initial conditions characterized by $r_a\geq r_{ac}(\alpha)$. Moreover, we also explored the behaviour of a few additional models with $2.5\leq \alpha\leq 2.9$ that confirmed the trends shown in Tab. \ref{tab1}. In practice for each value of $\alpha$, we started with models corresponding to $r_a=r_{ac}$ (the violently unstable models described in Sect. 3) and we systematically increased the value of $r_a$ up to a fiducial value for which the model remains stable over the simulations time $40t_*$. We then refined the determination of the critical value of $r_a$ by bisection around the successive approximations of$r_{as}$, and the values reported in Tab. \ref{tab1} give the numerical value of $r_{as}$ within $\pm 0.01$: for example for $\alpha=2$ the true value of $r_{as}$ is numerically found between 0.81 and 0.83.\\
\indent Figure \ref{figxistab} (top right) shows $r_{as}$ as a function of $\alpha$. We find that $r_{as}$ increases monotonically for increasing $\alpha$, showing however a curious plateau in the range $1.4\lesssim\alpha\lesssim 2$. The trend of $\xi_s$ with $\alpha$ shown in Fig. \ref{figxistab} (bottom right) and the corresponding values are given in Tab. \ref{tab1}. The behaviour of $\xi_s$ nicely mirrors that of $r_{as}$, with $\xi_s$ decreasing for increasing $\alpha$. Again, the plateau is clearly visible. For the Newtonian case ($\alpha=2$), we recover the well known result $\xi_s\simeq 1.7$ (corresponding to $r_{as}\simeq 0.8r_*$, see e.g. the review by \citealt{2011TTSP...40..425M} and references therein).\\ 
\indent Therefore, we conclude that equilibrium configurations in presence of forces with low values of $\alpha$ (i.e. forces ``longer-ranged" than Newtonian gravity) are able to support a larger amount of kinetic energy stored in radial-orbits than systems with larger force exponent $\alpha$. This trend of a greater stability for degreasing $\alpha$ is consistent with the special nature of systems with particles interacting with the harmonic oscillator force ($\alpha=-1$) where instabilities are impossible (\citealt{2004JSP...117..199L}, DCCN13). As it is well known when $\alpha=-1$ each particle, independently of the position of the others, oscillates in a time-independent harmonic oscillator field produced by a fixed point mass placed at the barycenter of the system and with a mass equal to the total mass of the system. It follows that no instabilities can develop. We interpret the numerical results of our simulations for decreasing $\alpha$ as the natural trend towards this behaviour. It is reasonable to expect that, when entering in the super harmonic regime ($\alpha<-1$, not explored in this paper),  collective phenomena will take
place again, similarly to what found in our collapse simulations (DCCN13).\\
\indent In line with these results we recall that NCL11 also found a higher value of $\xi_s$ ($\simeq 2.32$) for the OM Hernquist model in MOND than for its Newtonian counterpart without Dark Matter ($\simeq 1.64$). As can be seen from Tab. \ref{tab1}, $\xi_s$ for the $\alpha=1$ force (qualitatively similar to the deep MOND case) is larger than for $\alpha=2$. However, the analogy between additive $\alpha=1$ force and the deep MOND case is not (as expected) complete. In fact NCL11 also found, at variance with the present results for $\alpha=1$, that $r_{as}$ for the OM radially anisotropic Hernquist model in MOND is {\it larger} than $r_{as}$ for the corresponding one-component Newtonian system without Dark Matter. This different behaviour is due to the fact that the internal distribution of radial orbits is different in the two force laws, so a MOND model with larger $r_a$ than a Newtonian model does not necessarily have a lower $\xi$.\\
\indent We note that it has been suggested that the ROI in Newtonian systems is triggered by particles with orbital frequencies close to satisfying the condition $\Omega_P\equiv 2\Omega_\nu-\Omega_r\simeq 0$,
where $\Omega_\nu$ is the azimuthal frequency, $\Omega_r$ the radial
frequency and $\Omega_P$ the precession frequency (\citealt{1987MNRAS.224.1043P,1994ASSL..185.....P,1994LNP...433..143P}). Once a small non-spherical density perturbation is formed in a system dominated by low $\Omega_P$ orbits, it will grow more and more, as more and more particles tend to accumulate to it. In this interpretaion the time scale of the ROI therefore depends on the distribution of $\Omega_P$, and it is clear that models with fixed density profiles but different $r^{-\alpha}$ forces have different radial distributions of the precession frequencies. However this investigation is well beyond the scope of this paper.\\
\indent Following NCL11 and \cite{2015MNRAS.447...97G}, we also computed the value of the critical anisotropy parameter for stability when restricting to particles within the half-mass radius $r_*$. We call this quantity $\xi_{{\rm half},s}$, (see Fig. \ref{figxistab}  and Tab. \ref{tab1}). Not unexpectedly, the value of $\xi_{{\rm half},s}$ changes with $\alpha$ much less than $\xi_s$, this is a consequence of the radial trend of OM anisotropy, as summarized after Eq. (\ref{augmented}). Such result hints that a big enough, almost isotropic core could in principle stabilize against ROI a model with substantial radial anisotropy in the external regions, as found by \cite{2006ApJ...637..717T} in case of cold dissipationless collapses.\\
\begin{figure*}
\includegraphics[width=0.8\textwidth]{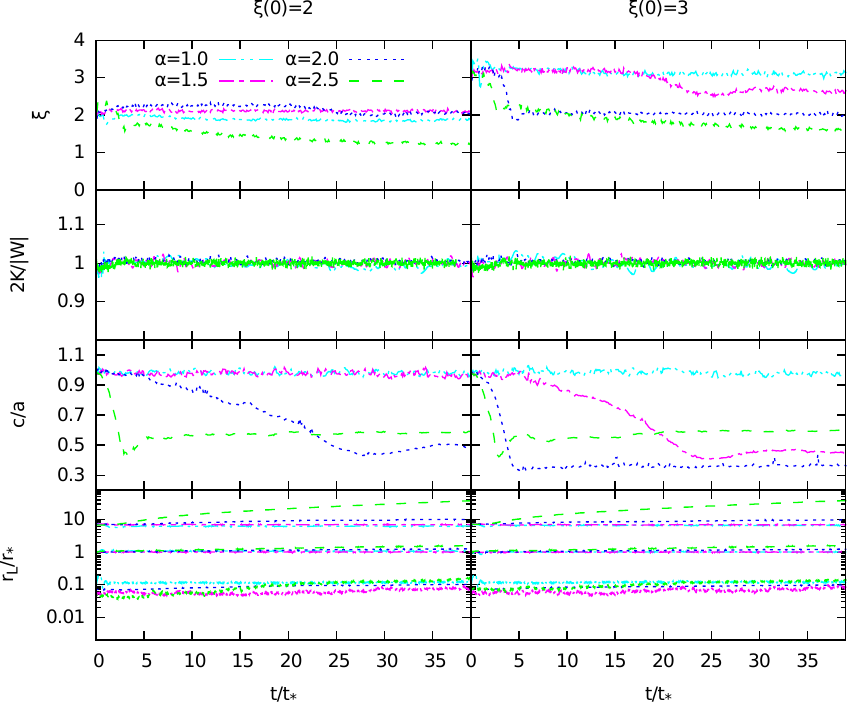}
\caption{From top to bottom: evolution of the stability indicator $\xi$, of the virial ratio $2K/|W|$, of the $c/a$ ratio and of the Lagrangian radii $r_5$, $r_{50}$ and $r_{90}$ for the anisotropic Hernquist models with $\alpha=1,$ 1.5, 2 and 2.5 and initial value of the anisotropy parameter $\xi(0)=2$ (left panels) and $\xi(0)=3$ (right panels). The stable cases for $\xi(0)=2$, and $\alpha=1$, 1.5 and for $\xi(0)=3$ and $\alpha=1$, are clearly visible.}
\label{figcaxi}
\end{figure*}
\begin{table}
\begin{tabular}{lccccccc}
\hline
$\alpha$ & $r_{ac}/r_*$ & $\xi_{c}$ & $\xi_{{\rm fin}}$  &     $(c/a)_{\rm fin}$        &   $r_{as}/r_*$ & $\xi_{s}$ & $\xi_{{\rm half},s}$\\
\hline
1.0      &  0.064       &  39.3     &      4.23          &        0.28                  &     0.58       &  3.87     & 1.59\\
1.1      &  0.065       &  25.6     &      3.77          &        0.27                  &     0.67       &  3.18     & 1.63\\
1.2      &  0.066       &  20.2     &      3.54          &        0.26                  & 0.78           &  2.76     & 1.49\\
1.3      &  0.067       &  17.6     &      3.29          &        0.27                  & 0.82           &  2.59     & 1.39\\
1.4      &  0.068       &  15.7     &      3.03          &        0.28                  & 0.85           &  2.15     & 1.33\\
1.5      &  0.069       &  13.5     &      2.87          &        0.28                  & 0.80           &  1.86     & 1.37\\
1.6      &  0.071       &  12.8     &      2.66          &        0.29                  & 0.83           &  1.83     & 1.34\\
1.7      &  0.074       &  11.4     &      2.52          &        0.31                  & 0.86           &  1.80     & 1.33\\
1.8      &  0.076       &  10.4     &      2.40          &        0.33                  & 0.83           &  1.78     & 1.35\\
1.9      &  0.078       &  8.95     &      2.23          &        0.36                  & 0.81           &  1.77     & 1.30\\
2.0      &  0.080       &  8.23     &      2.11          &        0.36                  & 0.82           &  1.71     & 1.31\\
2.1      &  0.084       &  6.70     &      1.99          &        0.40                  & 0.87           &  1.59     & 1.24\\
2.2      &  0.089       &  5.83     &      1.82          &        0.45                  & 0.91           &  1.44     & 1.22\\
2.3      &  0.100       &  4.64     &      1.75          &        0.47                  & 1.01           &  1.33     & 1.18\\
2.4      &  0.105       &  4.06     &      1.59          &        0.42                  & 1.10           &  1.27     & 1.15\\ 
2.5      &  0.125       &  3.44     &      1.33          &        0.56                  & 1.19           &  1.20     & 1.09\\
\hline
\end{tabular}
\caption{Properties of initial conditions and final states of the numerical models as functions of $\alpha$. From left to right columns: force exponent; normalized critical anisotropy radius for consistency; maximum value of the stability indicator; final value of the stability indicator; final value of the axial ratio for the maximally anisotropic models; normalized critical anisotropy radius for stability; corresponding value of the stability indicator; critical value of the stability indicator when restricting to particles within the half-mass radius $r_{50}=r_*$.}
\label{tab1}
\end{table}  
\begin{figure*}
\includegraphics[width=0.8\textwidth]{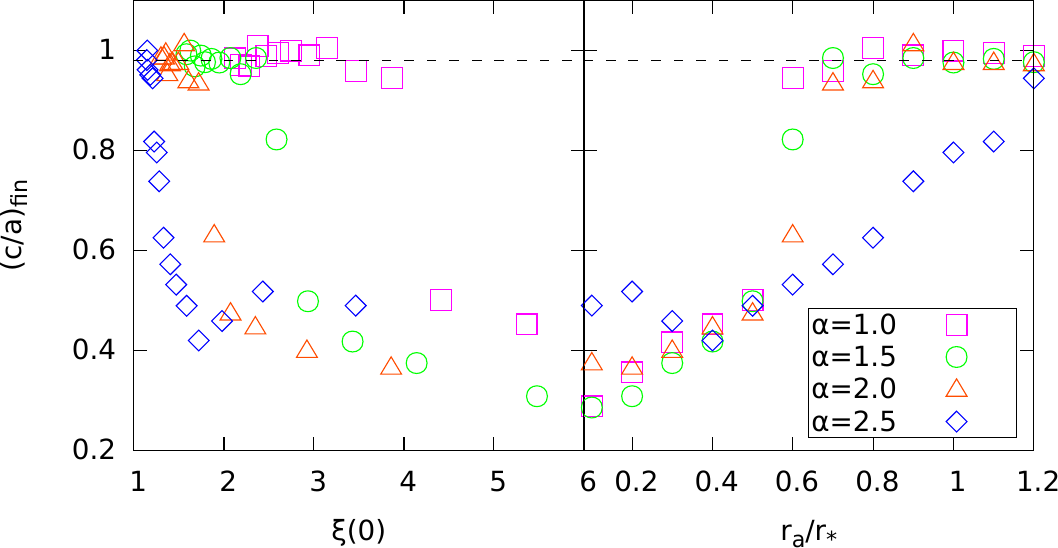}
\caption{Final axial ratios $(c/a)_{\rm fin}$ within $r_{90}$ versus the initial value of the stability indicator $\xi$ (left), and the normalized initial anisotropy radius $r_a/r_*$ (right) for $\alpha=1,$ 1.5, 2, and 2.5. The horizontal dashed line just marks the typical fiducial value of the final axial ratio $c/a$ for stability. It appears clear, from both indicators ($r_a$ and $\xi$), that the models with larger values of $\alpha$ are more prone to the ROI.}
\label{fig1}
\end{figure*}
\subsection{Evolution and end products of unstable models} 
Here we focus on the evolution of models with $\xi_s<\xi(0)<\xi_c$, which are unstable but not maximally anisotropic. As we have previously done for isotropic and maximally anisotropic models, for these unstable models we have monitored the evolution of the stability indicator $\xi$, of the virial ratio $2K/|W|$, of the axial ratio $c/a$, and of the three Lagrangian radii $r_5$, $r_{50}$ and $r_{90}$. These quantities are plotted in Fig. \ref{figcaxi} as functions of time (up to $40t_*$) for models with initial anisotropy parameters $\xi(0)=2$ (left panels), and 3 (right panels), and $\alpha=1$, 1.5, 2 and 2.5. For all these models the virial ratio and the Lagrangian radii do not show any significant change at late times, with the exception of the models with $\alpha=2.5$ that show a systematic drift with time of $r_{90}$ due to escaping particles (see also Sect. \ref{s32}). From the evolution of $\xi$ and $c/a$ it is apparent that, at fixed $\xi(0)$, models with higher $\alpha$ take less time (in units of $t_*$) to develop the ROI. For the models shown in Fig. \ref{figcaxi} the time $t_{\rm ROI}$ at which the instability sets in (defined as the time when the system departs significantly from the spherical symmetry; see Sect. 2.3) is never found to exceed $\simeq 7t_*$. However, unstable models with lower values of $\xi(0)$ are characterized by larger values of $t_{\rm ROI}$, up to $20t_*$.\\ 
\indent Figure \ref{fig1} shows for four relevant values of $\alpha=1,$ 1.5, 2 and 2.5 the final axial ratio $(c/a)_{\rm fin}$ as a function of both $\xi(0)$ and $r_a/r_*$. For similar values of $\xi(0)$, the end products of models with larger $\alpha$ tend to be more flattened, while, as a general trend, for fixed $\alpha$ models with smaller $r_a/r_*$ lead to more flattened end products, consistently with the results of Newtonian simulations (see e.g. \citealt{1996MNRAS.280..700P,1997ApJ...490..136M,2002MNRAS.332..901N,2009ApJ...704..372B}; NCL11). Interestingly, such behaviour is reminiscent of the trend of $(c/a)_{\rm fin}$ with the initial virial ratio $2K/|W|$ found in DCCN13 for cold collapses with $r^{-\alpha}$ forces: the end states of systems with lower initial virial ratio are more flattened. The physical origin of the findings of DCCN13 and of the present work is the same: collapses with colder initial conditions are more dominated by radial-orbits and therefore more affected by the ROI.\\
We note that for the maximally anisotropic models the final values of the stability indicator $\xi_{\rm fin}$ are systematically larger than $\xi_s$, as observed also for the end products of collapses dominated by radial orbits (see e.g. \citealt{2005A&A...433...57T,2007LNP...729..297E}). The reason for this is that such models have rather flattened end states (therefore not described by OM distribution functions), and considerations on the maximum amount of anisotropy for stability for spherical models can not be applied to non-spherical systems.\\  
\indent As we have observed for the end states of maximally anisotropic models, the averaged final density profile of unstable models does not depart significantly from the initial density profile. Consistently, the differential energy distribution $n(E)$ shows little to no evolution for such unstable systems (data not shown here). 
\section{Summary and Conclusions}
In this paper, the follow-up of a preliminary investigation (DCCN15), we have studied the onset of radial-orbit instability (ROI) in equilibrium Hernquist models with Osipkov-Merritt radial anisotropy and additive inter-particle forces of the form $r^{-\alpha}$, with $1\leq\alpha<3$. The aim of this work is to elucidate how the main properties of the ROI, routinely studied in the $\alpha=2$ case, depend on the force exponent $\alpha$, a measure of the ``force range". For the simulations we adopted our well tested direct $N$-body code, already employed for the study of cold collapses of self-gravitating systems with $r^{-\alpha}$ forces (DCCN13); the anisotropic equilibrium initial conditions have been constructed with the numerical inversion described in DCCN15. The main results are the following:
\begin{itemize}
\item We confirmed that, for all the explored of $\alpha$, isotropic models are associated with a phase-space distribution function $f(E)$ monotonic in terms of energy, and are numerically stable (DCCN15). Therefore, it is tempting to conjecture that the \cite{1973dgsc.conf..139A} theorem (see e.g. \citealt{2008gady.book.....B}) may hold not only in Newtonian gravity but more generally for $r^{-\alpha}$ forces. 
\item Following DCCN15 we determined the minimum value of the anisotropy radius for phase-space consistency, $r_{ac}$ and the associated value of the stability indicator $\xi_c$, confirming the preliminary indications that $r_{ac}$ increases and $\xi_c$ decreases for increasing $\alpha$. All marginally consistent models (i.e. models with $\xi$ close to $\xi_c$) are violently unstable and are characterized by strongly triaxial end products. In general the end products are more triaxial for lower values of $\alpha$, with $(c/a)_{\rm fin}\simeq 0.3$ for $\alpha=1$ increasing to $(c/a)_{\rm fin}\simeq 0.56$ for $\alpha=2.5$. Remarkably, even the most triaxial end products are not more elongated than E7 systems, as already found for the end products of cold collapses with $r^{-\alpha}$ forces (DCCN13). 
\item With $N$-body simulations we determined the fiducial value $r_{as}$ of the (minimum) critical anisotropy radius for stability and we found that its value increases with $\alpha$. The corresponding critical value of the stability indicator $\xi_{s}$ decreases for increasing $\alpha$. For $\alpha=2$ we found $\xi_s\simeq 1.7$ in agreement with previous works in Newtonian gravity. These results indicate that systems with high values of $\alpha$ can support smaller amounts of radial anisotropy than systems with similar density profiles but ``longer-range" forces. This is also in agreement with the fact systems with $\alpha=-1$ (harmonic oscillator inter-particle force) can not develop instabilities nor relaxation phenomena.
\item For $\alpha=1$ (MOND-like force), $\xi_s$ is larger than the correspondent quantity for the same model in Newtonian gravity, similarly to what happens for the Osipkov-Merritt radially anisotropic Hernquist model in deep-MOND regime. However, the values of $r_{as}$ and $\xi_s$ in the $\alpha=1$ and deep-MOND cases are quite different due to the different orbital distributions of the two models in their external regions.
\item Mildly unstable models (i.e. models that are unstable but not as extreme as models with $\xi\approx\xi_c$) develop significant changes in $c/a$ and $\xi$ well within $15t_*$ and typically such systems relax in (say) $10t_*$. Not unexpectedly the end products of all unstable models are less flattened than E7. In general at fixed $\alpha$ the flattening of the end products directly correlates with the value of $\xi$ of the initial condition. Instead the the differential energy distribution $n(E)$ remains qualitatively unaffected by the relaxation process, with the only exception of models with high values of $\alpha$.
\end{itemize}
The results of the present investigation, and those of DCCN15, then lead us to conclude that radial-orbit instability is a quite universal feature of collisionless sytems with low global angular momentum, not restricted to the special nature of the $r^{-2}$ force law.
\section*{Acknowledgements}
The Referee is warmly thanked for his/her insightful comments and useful suggestions. L.C. and C.N. acknowledge financial support from PRIN MIUR 2010-2011, project ``The Chemical and Dynamical evolution of the Milky Way and the Local Group Galaxies", prot. 2010LY5N2T. P.F.D.C. was partially supported by the INFN project DYNSYSMATH2016.
\bibliographystyle{mnras}
\bibliography{biblio.bib}
\end{document}